\documentclass[runningheads]{llncs}
\usepackage[T1]{fontenc}
\usepackage{graphicx,verbatim}
\usepackage{amsmath,amssymb}
\usepackage{bbding}
\usepackage[hidelinks]{hyperref}

\usepackage{booktabs}
\usepackage[table]{xcolor}

\begin{document}
\title{Longitudinal 3D Foundation Modeling for Neoadjuvant Breast Cancer Response Prediction from Serial DCE-MRI}
\titlerunning{Longitudinal 3D Foundation Modeling for pCR}
%


\author{Fidel Omar Tito Cruz \and
Neda Ghafouri \and
Zengyan Wang \and
Pegah Khosravi \and
Yu Tian \and
Chen Chen}
\authorrunning{F. O. Tito Cruz et al.}
%
\institute{
University of Central Florida, Orlando, FL, USA\\
\email{
fi513841@ucf.edu,
neda.ghafouri@ucf.edu,
zengyan.wang@ucf.edu,
pegah.khosravi@ucf.edu,
yu.tian@ucf.edu,
chen.chen@ucf.edu
}
}

  
\maketitle              

\begin{abstract}
Pathologic complete response (pCR) is an important endpoint in neoadjuvant chemotherapy (NAC) for breast cancer, and predicting pCR from imaging during treatment could support treatment response assessment. Many existing imaging-based approaches rely on a single static timepoint, which fails to capture changes that occur during treatment. In this work, we present a longitudinal framework that combines a frozen 3D foundation encoder (Pillar-0) with our Temporal Dynamics Network (TDN) to predict treatment response from serial Dynamic Contrast-Enhanced (DCE) MRI acquired across four clinical timepoints from pre-treatment to pre-surgery. The TDN combines time-aware volumetric embeddings with clinical and treatment data to predict pCR. Evaluated on 982 patients from the combined I-SPY2 and ACRIN-6698 cohort, the proposed model achieves strong performance across all reported metrics when longitudinal 3D imaging is fused with clinical data (test AUROC: 73.6\%, balanced accuracy: 69.1\%). While clinical variables provide the strongest individual predictive signal, longitudinal 3D imaging contributes complementary information when fused with clinical data, improving pCR prediction. Our source code is available at: \url{https://github.com/omarftt/longitudinal_temporal_pillar}.


\end{abstract}
\section{Introduction}

Neoadjuvant chemotherapy (NAC) is recommended for selected patients with breast cancer, particularly those with inflammatory or locally advanced disease and those with high-risk triple-negative or HER2-positive tumors for whom response to therapy may guide subsequent treatment decisions~\cite{Korde2021}. Among these patients, pathologic complete response (pCR) is an important marker of treatment response and is associated with an improved prognosis at the individual-patient level. It is widely used as an endpoint in neoadjuvant trials, although its value as a surrogate for long-term outcomes may vary across breast cancer subtypes and treatment contexts~\cite{Research_2020}. Earlier prediction of pCR could help identify patients who are unlikely to benefit from their current treatment regimen and potentially avoid unnecessary toxicity.

Most imaging-based approaches for pCR prediction use a single MRI acquisition, typically the pre-treatment (T0) scan. However, treatment-related changes occur throughout the course of NAC and may not be reflected in one scan alone. The I-SPY2 study includes serial MRI acquisitions at four clinical timepoints from pre-treatment to pre-surgery, providing an opportunity to study how imaging changes during therapy relate to treatment response. Modeling these serial 3D volumes remains challenging because the images are high-dimensional, clinical cohorts are relatively limited in size, and scans are acquired at uneven intervals. In addition, recent 3D foundation models are generally designed to process individual scans independently rather than serial imaging.

Our objective is to study whether pretrained 3D foundation-model representations can be adapted to the longitudinal and multimodal structure of neoadjuvant breast cancer response prediction. We formulate pCR prediction as temporal learning from serial DCE-MRI with irregular timing and missing imaging visits, combined with clinical and treatment information. To address this problem, we build on Pillar-0~\cite{agrawal2025pillar}, a 3D medical foundation model pretrained on 155k volumetric scans, and introduce a Temporal Dynamics Network (TDN) that models frozen volumetric embeddings across available treatment timepoints.

\begin{itemize}

\item \textbf{Longitudinal 3D foundation-model adaptation.}
We adapt frozen Pillar-0 representations from serial DCE-MRI for longitudinal pCR prediction, rather than treating the task as single-scan classification.

\item \textbf{Temporal modeling of serial MRI.}
We introduce a temporal module that uses scan timing and timepoint-availability masks to handle irregular and missing MRI visits.

\item \textbf{Multimodal prediction with a clinical prior.}
We combine longitudinal imaging with clinical and treatment variables, allowing imaging features to refine a fixed clinical prior.

\item \textbf{Evaluation on a combined breast MRI cohort.} 
We evaluate the framework on the combined I-SPY2/ACRIN-6698 cohort against clinical-only, 2D imaging, and single-timepoint 3D imaging baselines.

\end{itemize}

\section{Related Work}

\noindent\textbf{Foundation models for 3D volumetric representations.}
Large-scale vision foundation models have recently been applied to volumetric medical imaging. Pillar-0~\cite{agrawal2025pillar} is pretrained on 155k volumetric scans and reports strong performance on single-scan volumetric analysis. Other 3D foundation models, including Decipher-MR~\cite{yang2026deciphermrvisionlanguagefoundationmodel} and RadFM~\cite{Wu2025}, also learn radiology representations from large volumetric datasets. However, these models are primarily designed to process individual scans and do not directly address serial imaging acquired during treatment.

\vspace{0.35em}
\noindent\textbf{Temporal modeling of serial imaging.}
Neoadjuvant trials provide repeated MRI acquisitions during treatment together with clinically meaningful response labels. Many methods use this serial information for pCR prediction. BSTNet~\cite{Huang2026} uses self-supervised pretraining to learn from variable-length imaging sequences and considers differences in scan timing and molecular subtype. The Spatiotemporal Interaction model~\cite{tang2025sti} combines pre-treatment and on-treatment MRI for early pCR prediction. Janíčková et
al.~\cite{JanIva_Temporal_MICCAI2025} learn temporal representations from longitudinal breast MRI with a multi-task objective and predict pCR with a linear classifier on the learned features. Additionally, longitudinal medical imaging has been studied with recurrent~\cite{cui2019rnn_longitudinal_ad}, convolutional~\cite{zhang2020spatiotemporal}, and Transformer-based models~\cite{Holste2024}. BioViL-T~\cite{bannur2023learning} incorporates temporal image pairs into vision-language pretraining, but operates on 2D images rather than volumetric sequences.

These directions have largely been studied separately: most 3D foundation models focus on using individual scans, whereas temporal pCR models are typically task-specific networks. Our work combines frozen 3D foundation-model representations with serial DCE-MRI and patient clinical and treatment data in a multimodal pCR prediction framework.

\section{Method}

\begin{figure*}[!t]
\begin{center}
\includegraphics[width=0.95\linewidth]{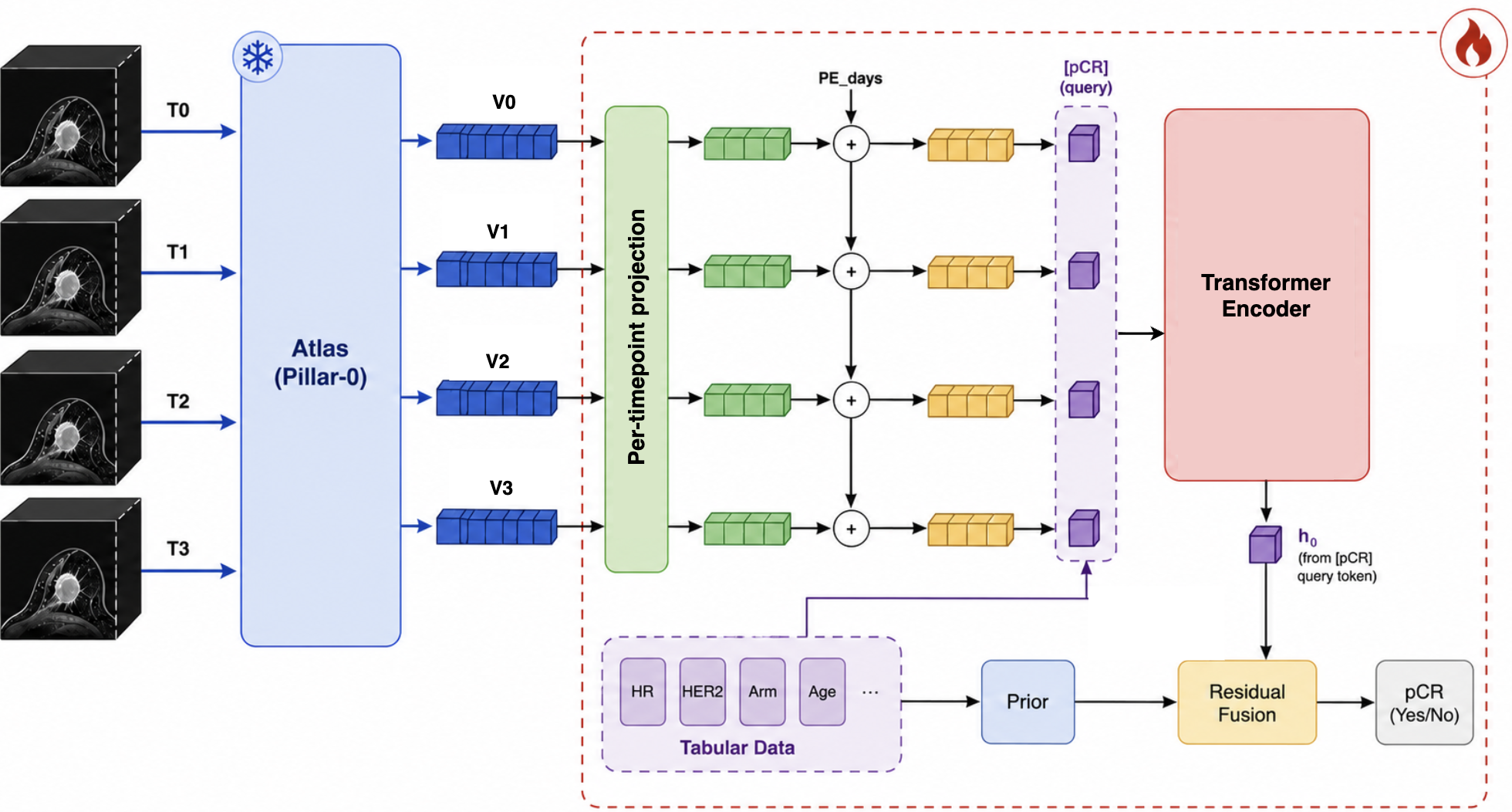}
\end{center}
\caption{End-to-end computational pipeline for longitudinal Pathologic Complete Response (pCR) prediction using 3D foundation model embeddings.}
\label{fig:implementation}
\end{figure*}

Our method extends Pillar-0 to the longitudinal setting by adding a temporal module over a patient’s serial imaging data. As shown in Fig.~\ref{fig:implementation}, each 3D MRI volume is independently encoded by a frozen vision backbone into a single embedding per timepoint. The temporal module projects these embeddings into compact imaging representations and incorporates an elapsed-time encoding measured from the pre-treatment scan. The resulting tokens, together with the patient’s clinical data, are processed by a Transformer encoder. 

\subsection{Visual Encoder}

For patient $i$ at timepoint $t \in {0,1,2,3}$, the input is a multi-channel 3D MRI volume $\mathbf{X}t^i \in \mathbb{R}^{C \times H \times W \times D}$, where $C{=}3$ corresponds to the pre-contrast, early post-contrast, and late post-contrast DCE phases, with $H{=}W{=}384$ and $D{=}192$. Following Pillar-0~\cite{agrawal2025pillar}, we use the pretrained 3D volumetric backbone $f{\text{vis}}$ to encode each volume independently. The backbone remains frozen throughout training:
\begin{equation}
\mathbf{v}_t^i = f_{\text{vis}}(\mathbf{X}_t^i;\,\theta_v)
\in \mathbb{R}^{1152}.
\end{equation}

\subsection{Temporal Dynamics Network}
\label{sec:tdn}

The TDN is the trainable component of the pipeline, shown as the shaded block in Fig.~\ref{fig:implementation}. It receives the frozen per-timepoint embeddings ${\mathbf{v}_0^i, \mathbf{v}_1^i, \mathbf{v}_2^i, \mathbf{v}_3^i}$ together with the patient’s clinical vector and predicts the pCR probability $\hat{y}^i$.

\vspace{0.35em}
\noindent\textbf{Per-timepoint projection.}
A projection module shared across timepoints transforms each high-dimensional volumetric embedding into a compact representation used for temporal modeling:
\begin{equation}
\mathbf{s}_t^i = \phi(\mathbf{v}_t^i) \in \mathbb{R}^{d},
\qquad d{=}64.
\end{equation}
The shared projection $\phi$ maps all timepoints into a common feature space, so their representations are comparable across serial scans.

\vspace{0.35em}
\noindent\textbf{Elapsed-time positional encoding.}
Serial scans are acquired at intervals that vary across patients, so we use the actual elapsed time since the pre-treatment scan rather than treating all scans as equally spaced. Let $a_t^i$ denote the number of days elapsed between the pre-treatment scan and timepoint $t$. We construct a temporal descriptor from linearly scaled and log-scaled elapsed-time terms together with the normalized scan order:
\begin{equation}
\boldsymbol{\rho}_t^i =
\Big[
\operatorname{clip}\big(a_t^i/180,\,0,\,2\big),\;
\log\big(1+a_t^i\big)/6,\;
t/(T{-}1)
\Big].
\end{equation}
We project the descriptor to $d$ dimensions and add it to the per-timepoint representation:
\begin{equation}
\mathbf{x}_t^i = \mathbf{s}_t^i + \psi(\boldsymbol{\rho}_t^i),
\qquad \psi(\boldsymbol{\rho}_t^i) \in \mathbb{R}^{d}.
\end{equation}
This allows the temporal model to account for irregular intervals between treatment scans while preserving their acquisition order.

\vspace{0.35em}
\noindent\textbf{Transformer encoder.}
We treat the time-aware per-timepoint representations as tokens and prepend a learnable query token $\mathbf{q}$. The clinical vector $\mathbf{c}^i \in \mathbb{R}^{17}$ (Sec.~\ref{sec:dataset}) is mapped by an MLP to a clinical token $\mathbf{x}_c^i$, which is appended to the token sequence. A Transformer encoder applies multi-head self-attention over
\begin{equation}
\left(
\mathbf{q},\,
\mathbf{x}_0^i,\,
\mathbf{x}_1^i,\,
\mathbf{x}_2^i,\,
\mathbf{x}_3^i,\,
\mathbf{x}_c^i
\right).
\end{equation}
A key-padding mask excludes unavailable timepoints from attention. The query token attends to the imaging and clinical tokens, and its output $\mathbf{h}^i \in \mathbb{R}^{d}$ serves as the patient representation. A linear layer maps this representation to a residual logit:
\begin{equation}
r^i = \mathbf{w}^{\top}\mathbf{h}^i + b.
\end{equation}

\vspace{0.35em}
\noindent\textbf{Residual fusion with a clinical prior.}
We combine the Transformer output with a clinical prior obtained from a logistic-regression model. The prior model $g_{\text{prior}}$ is fit on the clinical vector $\mathbf{c}^i$ using the training split:
\begin{equation}
\ell_{\text{prior}}^i = g_{\text{prior}}(\mathbf{c}^i).
\end{equation}
Its output is combined with the residual logit through a learned scalar weight $\alpha$:
\begin{equation}
\hat{y}^i =
\sigma\big(
\ell_{\text{prior}}^i + \alpha r^i
\big).
\end{equation}
The clinical prior provides a baseline prediction from the clinical variables, while the TDN learns a residual correction from the longitudinal imaging sequence and clinical context. The prior remains fixed after fitting.

\section{Experiments}

\subsection{Dataset and Clinical Features}
\label{sec:dataset}

We evaluate our method on the I-SPY 2 breast DCE-MRI cohort available through The Cancer Imaging Archive (TCIA)~\cite{li2020predicting,Newitt2021ACRIN6698}. We follow the cohort configuration and preprocessing organization provided by the curated BreastDCEDL-ISPY2 release~\cite{fridman2025breastdcedl_dataset}, which assembles 982 I-SPY2 trial patients distributed across two TCIA collections (717 under ISPY2 and 265 under ACRIN-6698/I-SPY2). The cohort includes serial DCE-MRI acquisitions at four planned clinical timepoints during NAC: pre-treatment (T0), early treatment (T1), inter-regimen (T2), and pre-surgery (T3). The I-SPY 2 trial acquired imaging prospectively at more than 20 clinical centers under a standardized protocol, on 1.5\,T and 3.0\,T scanners from three manufacturers (GE, Siemens and Philips) spanning 14 scanner models, and required all visits for a given patient to use the same scanner configuration, which limits scanner-related variability within a patient's series~\cite{li2020predicting,Newitt2021ACRIN6698}. The task is binary prediction of pCR, with an overall prevalence of approximately 32\%.

We use the predefined pCR-stratified split provided with BreastDCEDL-ISPY2 without modification. It contains 784 training, 99 validation, and 99 test patients, with a similar pCR prevalence across splits. The cohort includes patients with missing imaging timepoints, which are handled by excluding the corresponding tokens from attention through a key-padding mask.

Each patient is represented by a 17-dimensional clinical vector $\mathbf{c}^i \in \mathbb{R}^{17}$. It includes four binary indicators for hormone receptor status, HER2 status, MammaPrint risk, and menopausal status; age normalized by 100; and twelve binary indicators encoding the presence of individual drugs across treatment arms.

\subsection{Implementation and Evaluation}
\label{sec:implementation}

For each available timepoint, we construct a three-channel MRI volume from the pre-contrast, early post-contrast, and late post-contrast DCE phases. Each phase is resampled to $1\mathrm{mm}^3$ isotropic spacing, and the resulting volume is center-cropped or padded to $[3, 384, 384, 192]$, matching the Pillar-0 input specification. Intensities in each channel are clipped at the 1st and 99th percentiles and min-max scaled to $[0,1]$. Each volume is independently encoded by the frozen Pillar-0 backbone into a 1152-dimensional pooled embedding. For the 2D baselines, we use an ImageNet-pretrained ViT-MAE-base encoder on tumor-centered DCE-MRI slices. The T0 model is trained for pCR classification, while the same encoder is then frozen to extract per-timepoint embeddings for the longitudinal 2D baseline, which is processed by the same TDN and clinical-fusion pipeline.

Only the TDN and its prediction layers are trained. We use Adam with a learning rate of $5 \times 10^{-4}$, weight decay of $1 \times 10^{-4}$, and cosine annealing. Training uses a batch size of 64, dropout of 0.2, and gradient clipping. Models are trained for up to 150 epochs with early stopping based on validation AUROC. Class imbalance is handled with weighted binary cross-entropy. The logistic-regression clinical prior is fit on the training split for each seed and remains fixed during TDN training.

We report AUROC as the primary metric, together with sensitivity, balanced accuracy, PR-AUC, and negative predictive value (NPV). Results are reported as mean and standard deviation across 10 random seeds.

\begin{table}[t]
\centering
\caption{pCR prediction on the I-SPY2 test split. Values are percentages (mean $\pm$ std over 10 seeds).}
\label{tab:main_comparison}
\setlength{\tabcolsep}{4pt}
\resizebox{\textwidth}{!}{%
\begin{tabular}{l|ccccc}
\toprule
\textbf{METHOD} & \textbf{AUROC $\uparrow$} & \textbf{SENS $\uparrow$} & \textbf{BACC $\uparrow$} & \textbf{PRAUC $\uparrow$} & \textbf{NPV $\uparrow$} \\
\midrule
Clinical only
& $68.8_{\pm0.0}$
& $71.9_{\pm0.0}$
& $64.3_{\pm0.0}$
& $48.1_{\pm0.0}$
& $80.9_{\pm0.0}$ \\

2D imaging ($T_0$)
& $64.4_{\pm0.4}$
& $40.0_{\pm1.3}$
& $58.4_{\pm0.8}$
& $45.9_{\pm1.4}$
& $72.8_{\pm0.4}$ \\

2D imaging ($T_0$--$T_3$) + clinical
& $69.5_{\pm0.3}$
& $46.9_{\pm3.1}$
& $60.8_{\pm1.3}$
& $52.0_{\pm0.4}$
& $74.6_{\pm0.9}$ \\

3D imaging ($T_0$)
& $55.5_{\pm1.6}$
& $33.4_{\pm17.1}$
& $53.3_{\pm2.4}$
& $39.4_{\pm2.1}$
& $70.3_{\pm3.2}$ \\

3D imaging ($T_0$--$T_3$) + clinical (ours)
& \cellcolor{gray!25}$\mathbf{73.6}_{\pm1.2}$
& \cellcolor{gray!25}$\mathbf{73.8}_{\pm6.7}$
& \cellcolor{gray!25}$\mathbf{69.1}_{\pm1.1}$
& \cellcolor{gray!25}$\mathbf{53.8}_{\pm1.2}$
& \cellcolor{gray!25}$\mathbf{83.9}_{\pm2.0}$ \\
\bottomrule
\end{tabular}%
}
\end{table}

\section{Results and Discussion}

\subsection{Main Predictive Performance}
We evaluated the proposed longitudinal 3D model against clinical-only, 2D imaging, and single-timepoint 3D imaging baselines on the held-out test split. As shown in Table~\ref{tab:main_comparison}, the proposed longitudinal 3D model with clinical fusion achieves the best overall performance across all reported metrics, with an AUROC of 73.6, sensitivity of 73.8, balanced accuracy of 69.1, PR-AUC of 53.8, and NPV of 83.9. Compared with the clinical-only baseline, the full model improves AUROC by 4.8 points and balanced accuracy by 4.8 points.

The clinical-only baseline remains strong, reaching 68.8 AUROC and 64.3 balanced accuracy. This indicates that molecular markers, treatment arm, and other clinical variables carry substantial predictive signal in this cohort. However, the improvement achieved by the full model suggests that longitudinal 3D imaging provides complementary information beyond the clinical variables.

The imaging baselines show that single-timepoint imaging alone is less reliable. The 2D imaging model using only T0 reaches 64.4 AUROC, while the frozen 3D imaging model using only T0 reaches 55.5 AUROC, the lowest AUROC. This suggests that a single pooled 3D foundation-model embedding from the pre-treatment scan is a weak predictor of pCR when used alone. In contrast, modeling 3D embeddings across T0-T3 together with clinical data increases performance to 73.6 AUROC. The 3D model surpasses the 2D model in the longitudinal multimodal setting (73.6 versus 69.5), whereas the 2D model is stronger when only the pre-treatment scan is used (64.4 versus 55.5). These results suggest that the frozen volumetric representation is most useful when serial scan information is integrated with clinical data.

\begin{table}[t]
\centering
\small
\caption{Effect of the number of MRI timepoints on pCR prediction. Values are reported as percentages (mean $\pm$ std over 10 seeds).}
\label{tab:timepoints}
\begin{tabular}{l|ccccc}
\toprule
\textbf{TIME POINTS} & \textbf{AUROC $\uparrow$} & \textbf{SENS $\uparrow$} & \textbf{BACC $\uparrow$} & \textbf{PRAUC $\uparrow$} & \textbf{NPV $\uparrow$} \\
\midrule
$T_0$            & $68.3_{\pm3.8}$ & $41.2_{\pm13.5}$ & $58.3_{\pm3.9}$ & $49.0_{\pm3.5}$ & $73.3_{\pm3.6}$ \\
$T_0$--$T_1$     & $65.0_{\pm1.7}$ & $38.8_{\pm7.4}$ & $55.6_{\pm3.0}$ & $45.8_{\pm3.0}$ & $71.3_{\pm1.9}$ \\
$T_0$--$T_2$     & $71.9_{\pm0.9}$ & $73.1_{\pm4.2}$ & $68.6_{\pm1.0}$ & $51.6_{\pm2.0}$ & $83.4_{\pm1.3}$ \\
\textbf{$T_0$--$T_3$} & \cellcolor{gray!25}$\mathbf{73.6}_{\pm1.2}$ & \cellcolor{gray!25}$\mathbf{73.8}_{\pm6.7}$ & \cellcolor{gray!25}$\mathbf{69.1}_{\pm1.1}$ & \cellcolor{gray!25}$\mathbf{53.8}_{\pm1.2}$ & \cellcolor{gray!25}$\mathbf{83.9}_{\pm2.0}$ \\
\bottomrule
\end{tabular}
\end{table}

\subsection{Effect of Temporal Depth}
Table~\ref{tab:timepoints} evaluates the effect of adding MRI timepoints while keeping the encoder, temporal model, and clinical features fixed. The T0 model reaches 68.3 AUROC, while adding the early-treatment T1 scan decreases performance to 65.0 AUROC. This may indicate that the early interval provides limited additional signal beyond baseline, although the pattern should be interpreted as cohort-specific. Performance improves after including T2, reaching 71.9 AUROC, and is highest when all four timepoints are used, reaching 73.6 AUROC. This pattern indicates that the longitudinal signal becomes more useful at the inter-regimen and pre-surgery stages. Also, performance does not increase monotonically with the number of timepoints. Instead, the results indicate that the specific combination of timepoints, rather than the number of scans alone, affects prediction.

The most pronounced change is observed in sensitivity. It remains 41.2\% and 38.8\% for T0 and T0--T1, respectively, then increases to approximately 73\% once T2 is included. The T0-T2 configuration recovers much of the improvement obtained with all four timepoints, whereas T0-T3 provides a further increase in AUROC and PR-AUC. This suggests that mid-treatment imaging is already informative for prediction, and later imaging provides additional predictive information. The two settings also differ in when the required information becomes available. Predictions using T0-T2 can be made at the inter-regimen visit, whereas the T0-T3 model additionally requires the pre-surgical scan, postponing prediction until the end of treatment.

\subsection{Ablation Study}

Table~\ref{tab:ablation} examines the contributions of clinical features and temporal modeling to the frozen imaging representation. Imaging alone reaches 55.5 AUROC. Adding clinical features increases performance to 62.2 AUROC, while adding temporal modeling instead increases it to 64.0 AUROC. Neither partial configuration reaches the 68.8 AUROC of the clinical-only baseline in Table~\ref{tab:main_comparison}. In particular, combining a static T0 embedding with clinical features does not improve over clinical information alone, indicating that a single imaging embedding adds limited value when temporal information is not modeled.

Combining temporal modeling and clinical features yields the best result, reaching 73.6 AUROC, a larger gain than either component contributes on its own. This indicates that the contribution of frozen 3D imaging representations depends on how information from serial scans is modeled and integrated with clinical data. Rather than serving as strong standalone predictors, the frozen embeddings become useful within the full longitudinal multimodal framework, where the temporal module can use information across timepoints and the residual design builds on the clinical prior.

\begin{table}[t]
\centering
\caption{Ablation of our model (test split). Values are percentages (mean $\pm$ std over 10 seeds).}
\label{tab:ablation}
\setlength{\tabcolsep}{4pt}
\resizebox{\textwidth}{!}{%
\begin{tabular}{l|ccccc}
\toprule
\textbf{CONFIGURATION} & \textbf{AUROC $\uparrow$} & \textbf{SENS $\uparrow$} & \textbf{BACC $\uparrow$} & \textbf{PRAUC $\uparrow$} & \textbf{NPV $\uparrow$} \\
\midrule
Imaging (T0)   & $55.5_{\pm1.6}$ & $33.4_{\pm17.1}$ & $53.3_{\pm2.4}$ & $39.4_{\pm2.1}$ & $70.3_{\pm3.2}$ \\
Imaging (T0) + Clinical & $62.2_{\pm2.1}$ & $30.9_{\pm9.6}$ & $55.8_{\pm2.1}$ & $43.0_{\pm1.7}$ & $71.1_{\pm1.4}$ \\
Imaging (T0–T3) + Temporal  & $64.0_{\pm2.4}$ & $61.9_{\pm26.2}$ & $59.3_{\pm4.0}$ & $42.6_{\pm3.2}$ & $77.9_{\pm5.5}$ \\
\midrule
\textbf{Imaging (T0–T3) + Temporal + Clinical (Ours)} & \cellcolor{gray!25}$\mathbf{73.6}_{\pm1.2}$ & \cellcolor{gray!25}$\mathbf{73.8}_{\pm6.7}$ & \cellcolor{gray!25}$\mathbf{69.1}_{\pm1.1}$ & \cellcolor{gray!25}$\mathbf{53.8}_{\pm1.2}$ & \cellcolor{gray!25}$\mathbf{83.9}_{\pm2.0}$ \\
\bottomrule
\end{tabular}%
}
\end{table}

\section{Conclusions}
We presented a longitudinal framework that combines frozen Pillar-0 representations from serial DCE-MRI with a lightweight temporal module and clinical data for pCR prediction. On the combined I-SPY2 and ACRIN-6698 cohort, the full longitudinal multimodal model achieved the best performance across all reported metrics, reaching an AUROC of 73.6 and improving over the clinical-only baseline. The results show that clinical variables provide a strong baseline, while frozen 3D imaging embeddings are weak when used as static single-timepoint predictors. The imaging contribution is visible when serial scans are modeled jointly and integrated with clinical data, with the full model outperforming either partial configuration. The temporal-depth analysis further shows that much of the improvement is reached by mid-treatment, with T2 achieving performance close to the full four-timepoint model and T3 providing additional gains.

\vspace{0.35em}
\noindent\textbf{Prospects of Application:} This framework could estimate pCR likelihood from serial DCE-MRI and clinical variables during neoadjuvant therapy, potentially supporting review of treatment response at the inter-regimen stage. Prospective external validation, calibration, and integration into clinical workflows are required before clinical use.

\vspace{0.35em}
\noindent\textbf{Acknowledgment.} This material is based upon work supported by the UCF Seed Funding program.

\medskip

    

\begin{credits}

\subsubsection{\discintname}
The authors declare no competing interests.

\end{credits}

%
%
%
\bibliographystyle{splncs04}
\bibliography{mybibliography}

\end{document}